\documentclass[letterpaper,journal]{IEEEtran}
\usepackage{cite}
\usepackage{amsmath,amsfonts,amssymb}
\usepackage{mathrsfs}
\usepackage{mathtools}
\usepackage{array}
\usepackage{bm}
\usepackage[table,dvipsnames]{xcolor}
\usepackage{booktabs,tabularx}
\usepackage{multirow}
\usepackage{graphicx}
\usepackage{epstopdf}
\usepackage{epsfig}
\usepackage{url}
\usepackage{hyperref}
\hypersetup{hidelinks,hypertexnames=false}
\usepackage{balance}
\usepackage{stfloats}
\usepackage{algorithm}
\usepackage{algpseudocode}
\usepackage{paralist}

\begin{document}

\title{6G Native AI and Channel Foundation Models}

\author{Shugong Xu, \textit{Fellow, IEEE}, Jun Jiang, and Yuan Gao
\thanks{Shugong Xu and Jun Jiang are with Xi'an Jiaotong-Liverpool University, Suzhou, China, email: shugong.xu@xjtlu.edu.cn, and Jun.Jiang25@student.xjtlu.edu.cn.}
\thanks{Yuan Gao is with the School of Communication and Information Engineering, Shanghai University, Shanghai, China, email: gaoyuansie@shu.edu.cn.}}

\maketitle

\begin{abstract}
The integration of artificial intelligence (AI) and wireless communications is widely regarded as a core objective of sixth-generation (6G) systems. However, both the meaning of native AI and the type of AI capability that should be embedded into future wireless systems remain open to interpretation. This paper discusses 6G native AI from a system-design perspective and argues that native AI should be co-designed, optimized, and deployed as an intrinsic component of the wireless system rather than as a removable post-deployment add-on. From this perspective, conventional task-specific supervised models are difficult to use as the main technical basis of native AI because they depend heavily on labeled data, generalize poorly across propagation conditions, and require fragmented designs for different channel-related tasks. Motivated by these limitations, we position channel foundation models (CFMs) as a channel-centric foundation-model paradigm for 6G native AI. We define the scope of CFMs, clarify their differences from task-specific wireless AI models and large language models, and summarize three pretraining families: generative, discriminative, and hybrid pretraining. We further discuss how CFMs may support physical-layer processing, radio access network intelligence, and integrated sensing and communications. Preliminary CSI-CLIP-based results are included as bounded evidence that CFM-style pretraining can improve positioning and beam prediction when task-specific labels are limited.
\end{abstract}

\begin{IEEEkeywords}
6G native AI, channel foundation models, channel state information, masked channel modeling, contrastive learning, integrated sensing and communications, self-supervised learning.
\end{IEEEkeywords}

\section{Introduction}

Artificial intelligence (AI) has been widely investigated for mobile communication systems, including channel extrapolation, channel estimation, positioning, beam management, and network optimization~\cite{gao2026ai,pan2025ai,li2025survey}. As the 6G vision increasingly treats AI as a fundamental system capability, native AI has become a key term in discussions on future wireless networks. Nevertheless, there is still no common answer to what should be considered native AI or what type of AI capability future 6G systems should rely on.

This paper takes a system-oriented view. We interpret native AI as an AI capability that is built into the wireless system from the beginning of system design. It should be co-designed with the architecture, optimized with the air interface and network functions, and delivered as a necessary part of the system. This differs from an external AI module that is attached after deployment to solve an isolated problem. If native AI is expected to become a core component of 6G, it cannot rely primarily on a collection of separate models, each trained for one task and one scenario.

The requirement becomes clear when considering the expected operating conditions of 6G. Future networks must support terrestrial, aerial, maritime, and satellite communications, while also operating under highly dynamic propagation, heterogeneous interference, diverse carrier frequencies, and different antenna configurations. The same system may need to support channel estimation, channel state information (CSI) feedback, beam management, positioning, and integrated sensing and communications (ISAC). A task-specific model trained for one environment may work well within its training distribution, but it is unlikely to provide the task adaptability and scenario generalization needed by system-level native AI. If every task and scenario requires a separate model, the network architecture becomes increasingly fragmented. If models cannot generalize across channel domains, performance may fluctuate when the deployment environment differs from the training data.

This motivates a channel-centric foundation-model direction. Inspired by the success of foundation models in language and vision, channel foundation models (CFMs) aim to learn reusable channel representations from large-scale heterogeneous channel data and adapt them to downstream wireless tasks with limited task-specific supervision. The key point is not merely to use a larger neural network. Rather, CFMs shift the design focus from isolated task optimization to reusable channel representation learning. Because many wireless tasks depend on the same underlying propagation environment, a pretrained channel representation may provide a common technical basis for multiple 6G native-AI functions.

Related papers, datasets, and open-source implementations are curated in the Awesome Channel Foundation Models repository.\footnote{Awesome Channel Foundation Models: \url{https://github.com/GREAT-ISAC/Awesome-Channel-Foundation-Models}}

This article connects the system requirements of 6G native AI with the technical role of channel-centric pretraining and uses preliminary CSI-CLIP results to ground the discussion. It explains why native AI requires task-adaptive and scenario-generalizable channel intelligence and how CFMs can serve as a technical path toward that requirement. The main contributions are as follows:
\begin{itemize}
    \item We clarify a system-level interpretation of 6G native AI and derive three requirements for AI models embedded in 6G systems: task adaptability, scenario generalization, and deployment-aware scalability.
    \item We revisit the evolution from single-task supervised learning to multi-task learning and foundation-model-style pretraining, and explain why conventional task-specific wireless AI is insufficient as the main basis of native AI.
    \item We define CFMs as channel-specialized foundation models and distinguish them from task-specific supervised models, large language models (LLMs), and generic pretrained models.
    \item We summarize generative, discriminative, and hybrid CFM pretraining strategies and discuss their possible roles in physical-layer processing, radio access networks (RANs), and ISAC.
    \item We report preliminary CSI-CLIP-based results for positioning and beam prediction as bounded evidence for the value of channel pretraining under limited downstream supervision.
\end{itemize}

\section{Native AI Requirements in 6G}

\subsection{Native AI as a Built-in Capability}

Native AI should be understood as an inherent AI capability of the wireless system. In this sense, AI is not an external function added after a system has already been designed. Instead, it is included in the initial design space, including data collection, model updating, signaling overhead, inference latency, deployment location, and interactions with protocol functions. Such a definition changes the evaluation criterion for wireless AI. A model is not sufficient merely because it performs well on a single benchmark; it must also be reusable, maintainable, and compatible with system constraints.

The first requirement is task adaptability. A 6G network is expected to support communication, sensing, localization, and control-related functions. These tasks differ in their labels and objectives, but many of them are driven by the same channel state, multipath structure, mobility pattern, and blockage condition. A native-AI model should therefore reuse channel knowledge across tasks rather than rebuild independent models from scratch.

The second requirement is scenario generalization. 6G systems may span urban macro cells, indoor hotspots, industrial scenarios, unmanned aerial vehicle links, high-speed mobility, and satellite-terrestrial integration. The propagation distribution can vary significantly across these conditions. A model designed as a native capability should maintain useful representations under such distribution shifts, or at least support efficient adaptation with limited new data.

The third requirement is deployment-aware scalability. Native AI is not only an offline training problem. The model must fit storage, latency, energy, and signaling constraints. Very large models may be difficult to deploy at the edge or inside latency-sensitive radio functions. Conversely, very small task-specific models may be efficient but fail to generalize. A practical native-AI solution must balance representation capacity and deployment cost.

\subsection{Limitations of Task-Specific Wireless AI}

AI-enabled wireless communication has developed largely in parallel with the evolution of machine learning. The first stage is single-task supervised learning, where one dataset is used to train one model for one task. Examples include task-specific models for channel estimation, beam prediction, positioning, or channel feedback. These models can be effective in fixed settings, but their performance usually depends on the quality and quantity of labeled samples.

This dependence on labeled data is a severe limitation in wireless systems. Channel labels may require controlled measurements, ray-tracing simulation, environmental annotation, accurate positioning devices, or extensive calibration. In dynamic propagation environments, the distribution of interference, blockage, mobility, and user density can change over time. Stable and representative labeled datasets are therefore expensive to obtain and difficult to maintain.

The second limitation is weak generalization. A model trained on one propagation environment, frequency band, or antenna configuration may overfit the training distribution. When it is transferred to another scenario, its accuracy can degrade sharply. This is especially problematic for 6G, where terrestrial, aerial, maritime, and satellite domains may coexist and where the network may encounter channel conditions not covered by the training set.

The third limitation is task fragmentation. In a conventional design, each new wireless task often requires a dedicated architecture, feature pipeline, and training procedure. As the number of AI-supported functions grows, the system may contain a large number of specialized models. This increases storage cost, maintenance complexity, and integration difficulty, which is inconsistent with the idea of native AI as a coherent system capability.

Finally, many task-specific models have limited online adaptation capability. Models trained under static assumptions may not respond quickly to changes caused by mobility, obstruction, hardware variation, or environmental dynamics. This is a concern for 6G applications that require low-latency and reliable adaptation, such as autonomous driving, industrial control, and remote healthcare.

\section{From Wireless AI Models to CFMs}

\subsection{Evolution of Wireless AI Paradigms}

\begin{figure*}[t]
    \centering
    \includegraphics[width=0.92\textwidth]{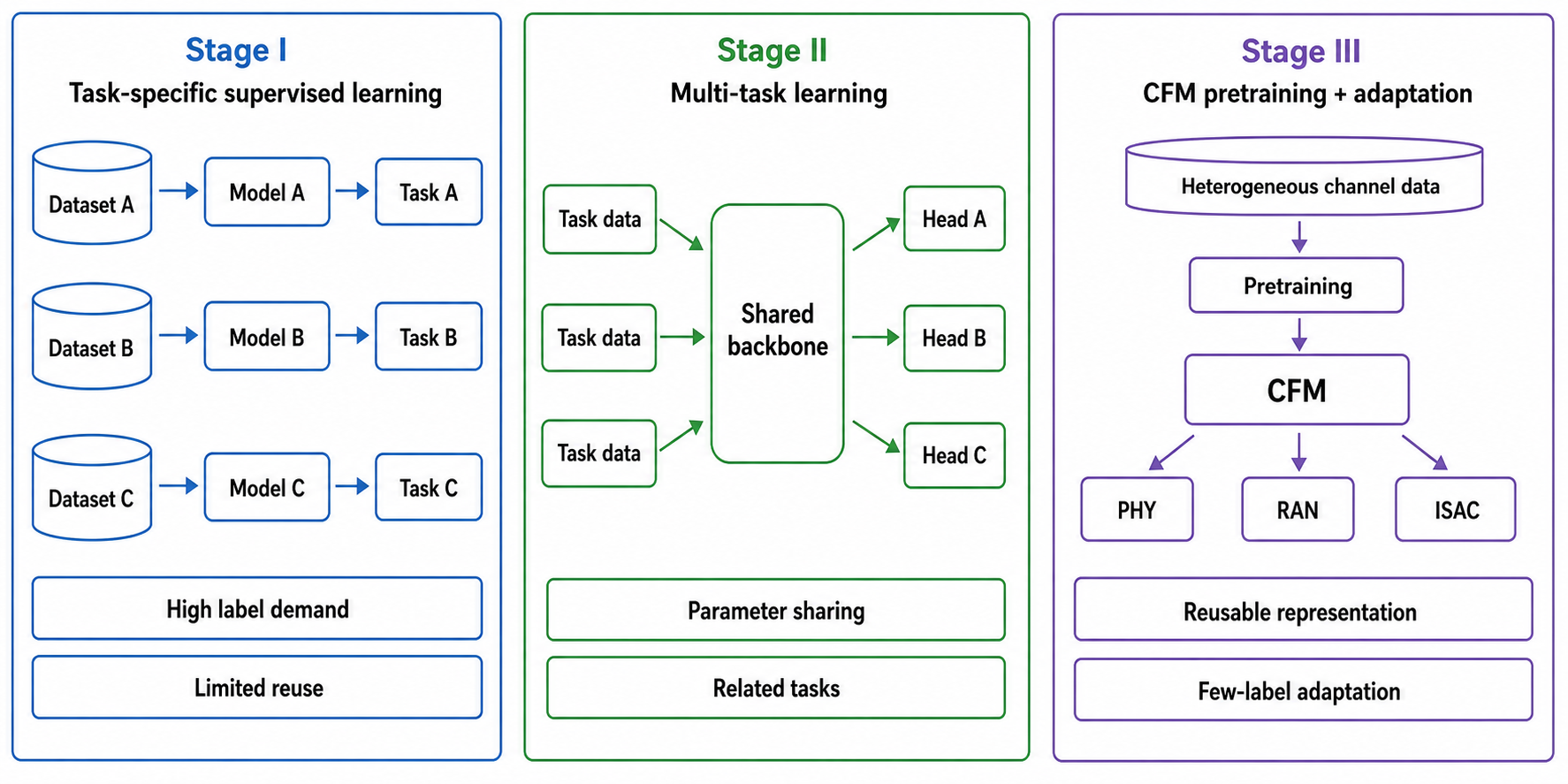}
    \caption{Evolution of AI paradigms for wireless communications. Single-task supervised learning optimizes isolated models, multi-task learning shares parameters across related tasks, and foundation-model-style pretraining learns reusable representations before downstream adaptation.}
    \label{fig:ai_paradigms}
\end{figure*}

The evolution of AI methods for wireless communications can be summarized in three stages, as illustrated in Fig.~\ref{fig:ai_paradigms}. The first stage is single-task supervised learning. It follows a ``one task, one dataset, one model'' paradigm and usually trains a neural network end to end for a specific wireless function. This paradigm is simple and can achieve strong performance under matched data distributions, but it is label-hungry and naturally limited in cross-task transfer.

The second stage is multi-task learning. By sharing part of the model across related tasks, multi-task learning can exploit common structures and reduce duplicated training effort~\cite{wang2023super,jiang2025mtca}. However, this paradigm still requires the tasks to have compatible input-output structures or strong statistical relationships. When task objectives are very different, such as reconstructing CSI, selecting a beam, and estimating a user location, a single multi-task architecture may need substantial task-specific customization.

The third stage is foundation-model-style pretraining followed by downstream adaptation. A foundation model is first trained on large-scale data, often with self-supervised objectives, and is then adapted to downstream tasks with lightweight finetuning or task-specific heads. For wireless communications, this paradigm is attractive because large amounts of unlabeled channel data can be obtained from simulation, measurement, or network operation, while labels for specific tasks remain expensive. The goal is to learn channel representations that are not tied to a single downstream task.

\subsection{Definition and Scope of CFMs}

CFMs are specialized foundation models for wireless channels. Their central object is the wireless channel rather than natural language, images, or generic network logs. A CFM is pretrained offline on large-scale heterogeneous channel data and learns representations that capture reusable propagation characteristics. The data may include CSI, channel impulse response (CIR), baseband in-phase/quadrature (I/Q) signals, or other channel-related measurements, depending on the intended task scope.

The input space of a CFM should be defined around channel-related observations. For CSI-based CFMs, the input may include raw complex CSI, amplitude-phase representations, angle-delay features, or multi-antenna subcarrier matrices. For broader radio CFMs, I/Q streams or spectrograms may also be used. The output of the pretrained model is usually a latent representation that encodes channel structure, such as multipath behavior, spatial-frequency correlation, delay-domain sparsity, blockage condition, or environment-dependent propagation patterns.

The task space of CFMs is also channel-centered. It can include generative or reconstruction-oriented tasks such as channel estimation, channel extrapolation, CSI compression and feedback, and precoding support. It can also include perception-oriented tasks such as scenario classification, positioning, beam management, and sensing. The common assumption is that these tasks depend on the same underlying channel physics and can benefit from a shared channel representation.

This definition distinguishes CFMs from three related concepts. First, CFMs are not simply task-specific supervised models with larger backbones. Their distinguishing feature is large-scale pretraining for transferable channel representations. Second, CFMs are not equivalent to LLMs. LLMs are pretrained mainly on language or general knowledge and may help network management or reasoning tasks, but they do not directly learn channel physics unless adapted with channel-specific data. Third, not every pretrained neural network in wireless communications is necessarily a CFM. A model pretrained for only one narrow task without reusable channel features is closer to a task-specific representation learner than to a channel foundation model.

\begin{figure}[t]
    \centering
    \includegraphics[width=0.98\columnwidth]{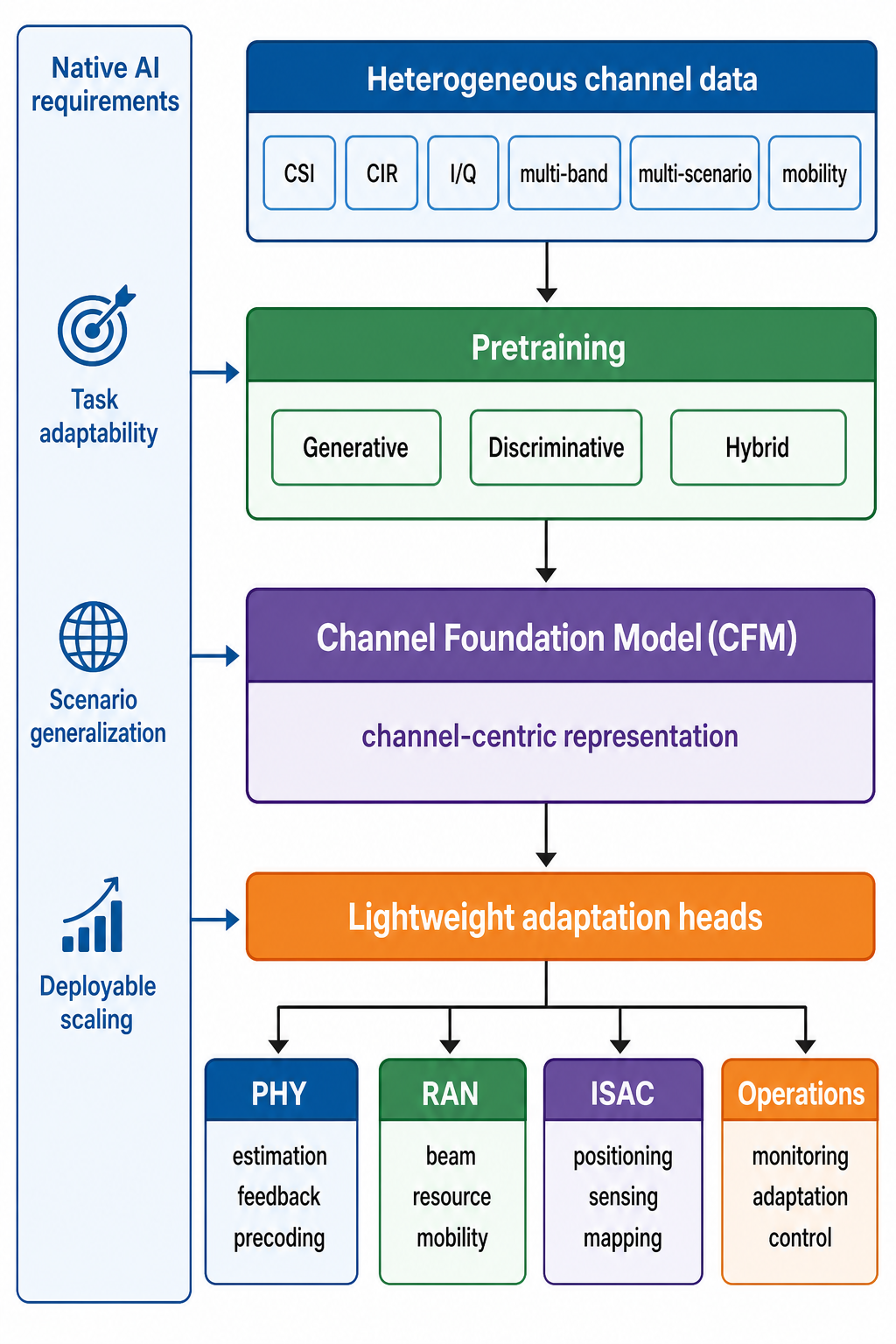}
    \caption{Conceptual framework of CFMs. A channel-oriented model is pretrained on heterogeneous wireless channel data and then adapted to downstream channel-related tasks through lightweight task-specific modules.}
    \label{fig:cfm_framework}
\end{figure}

\subsection{Comparison with Other AI Paradigms}

Table~\ref{tab:paradigm} compares task-specific supervised models, LLMs, and CFMs from the perspective of native AI. The table is qualitative and should be interpreted as a design comparison rather than a universal ranking. Task-specific models are attractive for low-latency deployment, but their task adaptability and scenario generalization are limited. LLMs provide language-interface and general-reasoning capabilities, but their large parameter scale and non-channel-centric pretraining make them less suitable as direct physical-layer channel models. CFMs aim to occupy the middle ground: they use channel-specific pretraining to improve adaptability and generalization while keeping the model scale more compatible with wireless deployment than general-purpose LLMs.

\begin{table}[t]
\centering
\caption{Qualitative Comparison of Wireless AI Paradigms}
\label{tab:paradigm}
\resizebox{\columnwidth}{!}{%
\begin{tabular}{lccc}
\toprule
Metric & Task-Specific Models & LLMs & CFMs \\
\midrule
Scenario generalization & Low & Medium & Potentially high \\
Task adaptability & Low & Medium & Potentially high \\
Storage overhead & Low & High & Medium \\
Parameter scale & Small & Very large & Medium \\
Pretraining requirement & No & Optional & Required \\
Inference latency & Low & High & Medium \\
Channel-physics alignment & Task-dependent & Indirect & Direct \\
\bottomrule
\end{tabular}}
\end{table}

\subsection{Core Properties of CFMs}

The first core property is cross-scenario and cross-configuration generalization. Traditional wireless AI models are often trained on narrow datasets and optimized for one scenario. CFMs instead rely on large-scale heterogeneous pretraining data that may cover different environments, frequency bands, antenna configurations, user distributions, and mobility patterns. By observing diverse channel realizations during pretraining, a CFM may learn more general channel statistics and reduce the need to relearn basic propagation features in each new deployment.

The second property is downstream-task adaptability. CFMs act as reusable channel feature extractors. Once the channel representation is pretrained, different task heads can be attached for channel estimation, beam prediction, positioning, channel identification, or sensing. This does not eliminate the need for task-specific data, but it reduces the amount of labeled data and model redesign required for each task.

The third property is scalability. In foundation-model research, scaling laws suggest that representation quality can improve with model capacity and data size under suitable training objectives. For CFMs, scalability has two dimensions. Model scalability refers to increasing model capacity to capture fine-grained channel dynamics, including nonlinear interactions among multipath components, interference, and antenna-domain correlations. Data scalability refers to training on broader channel data that cover more propagation environments and operating conditions. Both dimensions are important, but their practical value depends on whether the resulting model remains deployable under wireless latency and resource constraints.

\section{Pretraining Strategies for CFMs}

Pretraining is the mechanism that distinguishes a CFM from a larger task-specific wireless AI model. The objective should be chosen according to the channel structure that must be preserved and the downstream adaptation expected after pretraining. Existing CFM-oriented studies can be organized into three compact families.

\begin{itemize}
    \item \textit{Generative pretraining} reconstructs or predicts missing channel observations, such as masked CSI, CIR, or radio representations. This family naturally exploits unlabeled channel data and can learn correlations across antennas, subcarriers, time, and delay. WiFo and WirelessGPT are representative examples of reconstruction-oriented wireless foundation models~\cite{liu2025wifo,yang2025wirelessgpt}. Its main limitation is that low reconstruction loss does not automatically imply transferable channel knowledge, especially when local interpolation is sufficient to fill missing channel patches.
    \item \textit{Discriminative pretraining} shapes the representation space by distinguishing related and unrelated channel samples. CSI-CLIP is an example: it aligns frequency-domain CSI and delay-domain CIR because they describe the same propagation process in two domains~\cite{jiang2025mimo}. This objective encourages channel-consistent features, but its effectiveness depends on physically meaningful positive and negative pairs rather than arbitrary data augmentations.
    \item \textit{Hybrid pretraining} combines reconstruction and contrastive alignment. Such designs can preserve local channel structure while improving representation separation, as suggested by recent localization and channel-representation models~\cite{pan2025large,guler2026multi}. The practical challenge is to balance the losses so that one objective does not dominate the representation and reduce downstream transferability.
\end{itemize}

Across these families, the evaluation target should be downstream transfer rather than pretraining loss alone. A useful CFM objective should improve adaptation under limited labels, domain shifts, or new wireless tasks while remaining compatible with deployment constraints.

\section{CFMs for 6G Native AI}

\subsection{Physical-Layer Intelligence}

At the physical layer, CFMs may support channel estimation, channel feedback, channel extrapolation, and precoding-related tasks. Conventional methods often require task-specific training data and may degrade when the channel distribution changes. By contrast, a CFM pretrained on heterogeneous channel data can provide a reusable feature representation for these tasks. When only limited labeled data are available in a target scenario, lightweight finetuning can adapt the pretrained representation without retraining the entire model from scratch.

This does not mean that CFMs replace model-based signal processing. Instead, CFMs should be viewed as a representation-learning component that can be combined with wireless priors. For example, channel sparsity, antenna geometry, delay-Doppler structure, and pilot design can guide the input representation and pretraining objective. Such physics-aware design is important for avoiding black-box models that work only under narrow simulated conditions.

\subsection{Radio Access Network Intelligence}

The radio access network (RAN) connects user equipment with the core network and is responsible for resource scheduling, interference management, mobility control, and beam management. CFMs may provide a channel-aware representation for these functions. In dense urban deployments, beam selection is affected by blockage, user movement, cell-sector geometry, and inter-cell interference. A task-specific supervised beam predictor may work in one deployment but require retraining in another. A pretrained channel representation can reduce this dependence by transferring propagation knowledge across scenarios.

For native AI, this matters because RAN intelligence should not become a collection of unrelated models. If beam management, positioning assistance, and channel condition classification share a common channel representation, the network can reduce duplicated training and simplify deployment. However, practical use still requires careful evaluation of latency, signaling overhead, model update frequency, and robustness to out-of-distribution channels.

\subsection{Integrated Sensing and Communications}

ISAC introduces another requirement for native AI: the same wireless signals may support both communication and sensing. Communication objectives often focus on reliable data transmission, while sensing objectives may focus on localization, tracking, imaging, or environment understanding. These objectives are different, but they are linked through the same propagation environment.

CFMs can help by learning a unified representation of channel features that are useful for both communication and sensing. For example, multipath structure, delay spread, angular information, and blockage patterns can support both beam management and positioning. A CFM-based ISAC pipeline may therefore reduce the gap between communication-oriented and sensing-oriented processing. This is particularly relevant for intelligent transportation, indoor localization, and environment-aware networks, where spatial information and communication performance are tightly coupled.

\subsection{Preliminary Evidence from CSI-CLIP}

To provide bounded evidence for the CFM concept, we report CSI-CLIP-based experiments using a Vision Transformer (ViT) encoder. The pretraining data are constructed from the DeepMIMO dataset and contain more than 700,000 samples across 35 representative wireless scenarios. The scenarios cover indoor and outdoor environments and span sub-6 GHz, millimeter-wave, and terahertz frequency ranges. A non-pretrained ViT model is used as the baseline.

Table~\ref{tab:csiclip} summarizes the transfer results. For positioning, CSI-CLIP reduces the error in all nine city scenarios and achieves an average relative improvement of 21.57\% over the non-pretrained ViT baseline. For beam prediction, it improves accuracy in all six scenarios, with gains from 1.75 to 2.78 percentage points. These results support the bounded claim that channel pretraining can reduce dependence on task-specific labels and improve transfer for related downstream tasks. The boundary is also clear: the evidence is based on CSI-CLIP and DeepMIMO-derived scenarios, so it should not be interpreted as universal proof for all CFMs or all measurement environments.

\begin{table*}[t]
\centering
\caption{Preliminary CSI-CLIP Evidence Under Limited Downstream Supervision}
\label{tab:csiclip}
\resizebox{\textwidth}{!}{%
\begin{tabular}{lllll}
\toprule
Downstream Task & Evaluation Scope & Baseline & CSI-CLIP Result & Main Observation \\
\midrule
Positioning & 9 city scenarios; 503--2764 finetuning samples & Non-pretrained ViT & Lower error in all scenarios & 21.57\% average relative improvement \\
Beam prediction & 6 scenarios & Non-pretrained ViT & Higher accuracy in all scenarios & 1.75--2.78 pp accuracy gain \\
\bottomrule
\end{tabular}}
\end{table*}

These experiments do not settle the CFM research problem. They indicate that channel pretraining can be useful when downstream labels are scarce, but future work must test broader measurement datasets, cross-device and cross-time splits, hardware impairments, pilot overhead, latency, model compression, and online adaptation. These factors are necessary before CFMs can be treated as deployable native-AI components.

\section{Conclusion}

This paper has discussed 6G native AI and channel foundation models from a system-design perspective. We argued that native AI should be embedded into 6G systems as an intrinsic capability and should therefore provide task adaptability, scenario generalization, and deployment-aware scalability. Conventional task-specific supervised models are difficult to use as the main basis of native AI because they depend heavily on labeled data, generalize poorly across propagation conditions, and require fragmented designs for different tasks. CFMs provide a possible technical option by learning reusable channel representations from large-scale heterogeneous channel data and adapting them to downstream wireless tasks. We defined the scope of CFMs, distinguished them from task-specific models and LLMs, summarized generative, discriminative, and hybrid pretraining strategies, and discussed their possible roles in the physical layer, RAN, and ISAC. Preliminary CSI-CLIP-based results support the potential of this direction for positioning and beam prediction under limited downstream supervision, but broader experimental validation and deployment-aware design remain necessary for future 6G native-AI systems.

\bibliographystyle{ieeetr}
\bibliography{Myreference}

\end{document}